\documentclass[showpacs,floatfix]{revtex4}

\usepackage{amsmath}
\usepackage{graphicx}

\newcommand{\be}{\begin{equation}}
\newcommand{\ee}{\end{equation}}
\newcommand{\bea}{\begin{eqnarray}}
\newcommand{\eea}{\end{eqnarray}}
\newcommand{\bse}{\begin{subequations}}
\newcommand{\ese}{\end{subequations}}

\newcommand{\btheta}{\boldsymbol{\theta}}
\newcommand{\bxi}{\boldsymbol{\xi}}
\newcommand{\av}[1]{{\left\langle#1\right\rangle}}

\newcommand{\F}{{\mathcal F}}

\newcommand{\A}{\mathcal A}
\newcommand{\h}{\textsl{h}}

\begin{document}

\title{Searching for gravitational waves from known pulsars at once and twice the spin frequency}

\author{Micha{\l} Bejger}
\affiliation{N. Copernicus Astronomical Center, Bartycka 18, 00-716, Warszawa, Poland}
\email{bejger@camk.edu.pl}

\author{Andrzej Kr\'olak}
\affiliation{Institute of Mathematics, Polish Academy of Sciences, \'Sniadeckich 8,
00-956 Warszawa, Poland}
\email{krolak@impan.gov.pl}

\begin{abstract}
The existence of a superfluid core in the interior of a rotating
neutron star may have an influence on its gravitational wave emission.
In addition to the usually-assumed pure quadrupole radiation with the
gravitational wave frequency at twice the spin frequency, a frequency
of rotation itself may also be present in the gravitational wave spectrum.
We study the parameters of a general model for such emission, compare
it with previously proposed, simpler models, discuss
the feasibility of the recovery of the stellar parameters
and carry out the Monte Carlo simulations to test the performance
of our estimation method.
\end{abstract}


\maketitle

\section{Introduction}

Rotating, deformed neutron stars (NSs) are promising sources of gravitational
waves (GWs). They radiate GWs because of the non-vanishing, changing in time mass
quadrupole moment, i.e., non-axisymmetric mass distribution around the rotation
axis as seen by the distant observer. The departure from the
axisymmetric shape may be an outcome of the internal magnetic field and/or
elastic stresses in the crust/core and, if detected, will provide an
interesting insight into presently not very well known details of the
interior NS structure (for a recent review, see \cite{Andersson2011}).

The most commonly considered NS model used in the GW data analysis assumes a
triaxial star rotating about one of the principal axes of its moment of
inertia. In such a case one expects the GW frequency to be equal twice the
rotational frequency, $\Omega_{\rm GW}=2\Omega$. In a more general case, when
the axis of rotation is inclined w.r.t. the principal axis of the moment of
inertia (by an angle $\theta$, say), the resulting GW will be emitted at both
$\Omega$ and $2\Omega$ frequencies. Such a case was discussed by
\cite{ZimmermannS1979}, where it was assumed that the star is a rigid body. The
model's additional outcome is therefore a free precession of the spin axis
about the total angular momentum direction of the system. However, there is
currently no robust observational evidence of free precession in the population
of known NSs.  Secondly, the interior of the NS is most likely a fluid - the
rigid part (the crust) is only a small fraction of the total stellar mass (see,
e.g., \cite{HaenselPY2007}), and hence the precession frequency is expected
to be orders of magnitude smaller as compared to the rigid case
\cite{JonesA2001}. A model of a rotating, {\it completely fluid} star with the
mass quadrupole generated by distortional pressure from the star's magnetic
field was presented by \cite{BonazzolaG1996}, with the GW radiation at
$\Omega$ and $2\Omega$ frequencies.

Recently, D. I. Jones \cite{Jones2010} provided an important generalization of
the NS model of emission that takes into account the core superfluid component
`pinning' to the solid crust. There is now a compelling evidence that NSs
contains superfluid components in their interiors (see observations of
Cassiopeia A supernova remnant NS cooling and interpretation, e.g.,
\cite{ShterninYHHP2011}).  The model \cite{Jones2010} explores a possibility of
the superfluid `pinning' along an axis that is not one of the principal axes of
the star's moment of inertia, and concentrates on the allowed non-precessional
rotation of the star.  The resulting GW emission occurs at $\Omega$ and
$2\Omega$ frequencies.

\begin{figure}[h]
\begin{center}
\resizebox{0.4\columnwidth}{!}
{\includegraphics{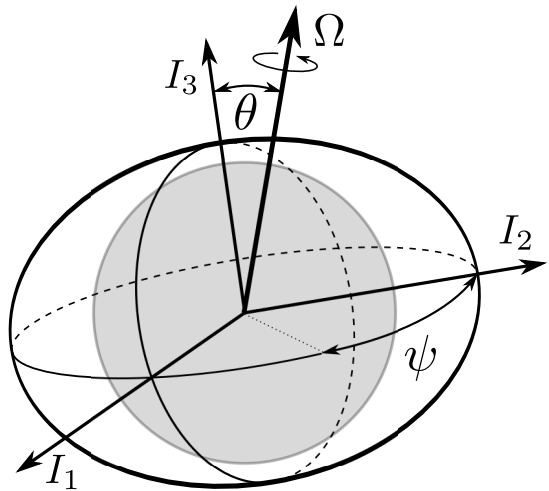}}
\end{center}
\caption{Triaxial neutron star rotating about a non-principal axis of inertia:
$\theta$ and $\psi$ are the orientation angles of the rotation axis of the superfluid component (gray region)
in the frame of the principal axes of the moment of inertia of the crust (for details
see \cite{Jones2010}).}
\label{fig:sf-angles}
\end{figure}

The above-mentioned models are summarized in Fig.~\ref{fig:sf-angles}.
Triaxial rigid star radiating at $2\Omega$ corresponds to $\theta\equiv 0$.
Assuming homogeneous interior, $I_1=I_2$ and the wobble angle $\theta\neq 0$
will result in a rigid, biaxial, precessing star of \cite{ZimmermannS1979} and
\cite{BonazzolaG1996}. Gray region denotes a spherical superfluid component
introduced in \cite{Jones2010}.

We study here the detection and parameter estimation of almost monochromatic
GWs emitted by known solitary pulsars in the data collected by a
detector. We thus assume that the frequency of the wave (together with its time
derivatives, i.e., the spindown parameters) and the position of the source in
the sky are known.  Several searches for known pulsars were already performed
with data collected by the LIGO, Virgo and GEO600 detectors
\cite{LSC04,LSC05,LSC07,LSC08,LSC10,LSC13}, assuming that pulsars emits GWs at
twice their spin frequency only. We consider the parameter estimation for models
where a pulsar emits the GWs at {\it both} once and twice its
spin frequency. The article is composed as follows: Sect.~\ref{sect:signal}
describes the response of the detector to the GW signal at both once
and twice the spin frequency as proposed by \cite{Jones2010}. Section~\ref{sect:stat} derives the
statistic for detection of the signal introduced in Sect.~\ref{sect:signal}
in white Gaussian noise with unknown variance. The maximum likelihood estimators
for parameters of the signal and the variance of the noise are also obtained.
Section~\ref{sect:snr} derives expressions for the signal-to-noise ratio for the
considered model, averaged over certain parameters, Sect.~\ref{sect:fisher} describes the Fisher matrix,
Sect.~\ref{sect:DataMethod} presents our parameter estimation method
and Sect.~\ref{sect:mcsim} contains results of the Monte Carlo simulations of the method for
different models of GW emission.
Section~\ref{sect:concl} contains conclusions.

\section{Gravitational wave signal at once and twice the rotation frequency
of a pulsar}
\label{sect:signal}
The template for the GW signal from a rotating superfluid NS (\cite{Jones2010})
depends on a set of the following parameters:
$\btheta=(h_0,h_1,\phi_o,\psi_o,\iota,\theta,\psi,\delta,\alpha,\mathbf{\omega})$,
where $h_0$ and $h_1$ are the dimensionless amplitudes,
$\phi_o$ is an initial phase,
$\psi_o$ and $\iota$ are the polarization and inclination angles (see e.g.,
\cite{NR04,NR08}), and
$\theta$ and $\psi$ are orientation angles of the superfluid component
in the frame of the principal axes of the moment of inertia of the crust.
Amplitudes $h_0$ and $h_1$ depend in the following way on the principal
moments of inertia $I_i, i = 1, 2, 3$:
\be
h_0 = \frac{4 \Omega^2 (I_3 - I_1)}{r},\quad h_1 = \frac{4 \Omega^2 (I_2 - I_1)}{r},
\ee
with $\Omega$ being the angular spin frequency of the pulsar and $r$
the distance to the pulsar. Angles $\delta$ (declination) and $\alpha$
(right ascension) are equatorial coordinates
determining the position of the source in the sky,
and the `frequency vector' $\mathbf{\omega}:=(\Omega,\Omega_1,\Omega_2,\dots)$
collects the frequency $\Omega$ and the spindown parameters (frequency derivatives) of the signal.
In the case of pulsars known from radio observations
the subset $\bxi=(\mathbf{\omega},\delta,\alpha)$ of the parameters $\btheta$
is assumed to be given.

The response $s(t)$ of an interferometric detector to the GW signal derived in \cite{Jones2010} is a sum of two
components $s_1(t)$ and $s_2(t)$ corresponding to GW frequencies of $\Omega$ and $2\Omega$.
The two components can be written in the following form:
\bea
\label{eq:sig1}
s_1(t) = \sum^4_{k = 1} A_{1k} h_{1k}(t),\\
\label{eq:sig2}
s_2(t) = \sum^4_{k = 1} A_{2k} h_{2k}(t),
\eea
where the eight functions of time $h_{lk}$ $(l=1,2, k=1,\ldots,4)$ depend only on parameters $\bxi$, and are defined as follows
\be
\label{eq:amps}
\begin{array}{ll}
h_{l1}(t;\bxi) := a(t;\delta,\alpha) \cos l\phi(t;\mathbf{\omega},\delta,\alpha),
&\quad
h_{l2}(t;\bxi) := b(t;\delta,\alpha) \cos l\phi(t;\mathbf{\omega},\delta,\alpha),
\\[1ex]
h_{l3}(t;\bxi) := a(t;\delta,\alpha) \sin l\phi(t;\mathbf{\omega},\delta,\alpha),
&\quad
h_{l4}(t;\bxi) := b(t;\delta,\alpha) \sin l\phi(t;\mathbf{\omega},\delta,\alpha),
\end{array}
\ee
with $a$, $b$ denoting the amplitude modulation functions, and $\phi$
the phase modulation function. Their explicit forms are given in \cite{JKS98}.
For the case of a pulsar known from radio observations the functions given by
Eqs.~(\ref{eq:amps}) are known.

In the model proposed by Jones \cite{Jones2010} the time independent amplitudes $A_{lk}$ depend
in general on 7 extrinsic parameters $(h_0,h_1,\phi_o,\psi_o,\iota,\theta,\psi)$.
However, it was recently indicated \cite{DIJ2012unp} that the model has 6 independent parameters only.
The independent parameters are the angles $\psi_o$ and $\iota$ that determine polarization
of the wave and 4 other parameters, $G_1, G_2, H_1, H_2$, that depend on the remaining 5 parameters
$(h_0, h_1,\iota,\theta,\psi)$:
\bea
&G_1 = k_1\,\cos\phi_o + k_2\,\sin\phi_o, \quad
\label{eq:G}
G_2 = k_1\,\sin\phi_o - k_2\,\cos\phi_o,\\
&H_1 = k_3\,\cos 2\phi_o + k_4\,\sin 2\phi_o, \quad
\label{eq:H}
H_2 = k_3\,\sin 2\phi_o - k_4\,\cos 2\phi_o,
\eea
where
\bea
k_1 =  \sin2\theta\,(h_0 - h_1\cos^2\psi)/2,&\quad&
k_2 =  h_1 \sin\theta\,\sin2\psi/2,\\
k_3 = -\Big(h_1(\cos^2\theta \cos^2\psi - \sin^2\psi) + h_0\sin^2\theta\Big),&\quad&
k_4 =  h_1\cos\theta\,\sin2\psi.
\eea
The 8 amplitude parameters are then given by
\be
\label{eq:amp8}
\begin{aligned}
A_{11} =  C_1 G_1 - C_2 G_2, &\quad A_{12} =  C_3 G_1 + C_4 G_2,\\
A_{13} = -C_1 G_2 - C_2 G_1, &\quad A_{14} = -C_3 G_2 + C_4 G_1,\\
A_{21} =  D_1 H_1 - D_2 H_2, &\quad A_{22} =  D_3 H_1 + D_4 H_2,\\
A_{23} = -D_1 H_2 - D_2 H_1, &\quad A_{24} = -D_3 H_2 + D_4 H_1,
\end{aligned}
\ee
with
\be
\begin{aligned}
&C_1 = A_{1+}\cos 2\psi_o,\quad C_2 = A_{1\times}\sin 2\psi_o,\quad C_3 = A_{1+}\sin 2\psi_o,\quad C_4 = A_{1\times}\cos 2\psi_o,\\
&D_1 = A_{2+}\cos 2\psi_o,\quad D_2 = A_{2\times}\sin 2\psi_o,\quad D_3 = A_{2+}\sin 2\psi_o,\quad D_4 = A_{2\times}\cos 2\psi_o,
\end{aligned}
\ee
and the constants $A_+$ and $A_\times$,
\be
\label{aa}
A_{1+} := \frac{1}{2} \sin\iota\cos\iota,\quad
A_{1\times} := \frac{1}{2} \sin\iota, \quad
A_{2+} := \frac{1}{2} (1 + \cos^2\iota),\quad
A_{2\times} := \cos\iota.
\ee

For $h_1 \equiv 0$ the amplitudes become independent of
the orientation angle $\psi$ and they depend on 5 parameters only:
\be
\label{eq:ampone}
\begin{array}{l}
A_{11} =  h_0 \sin2\theta \big(A_{1+}\cos2\psi_o\cos\phi_o - A_{1\times}\sin2\psi_o\sin\phi_o\big)/2,
\\[1ex]
A_{12} =  h_0 \sin2\theta \big(A_{1+}\sin2\psi_o\cos\phi_o + A_{1\times}\cos2\psi_o\sin\phi_o\big)/2,
\\[1ex]
A_{13} = -h_0 \sin2\theta \big(A_{1+}\cos2\psi_o\sin\phi_o - A_{1\times}\sin2\psi_o\cos\phi_o\big)/2,
\\[1ex]
A_{14} = -h_0 \sin2\theta \big(A_{1+}\sin2\psi_o\sin\phi_o + A_{1\times}\cos2\psi_o\cos\phi_o\big)/2,
\\[1ex]
A_{21} =  h_0 \sin^2\theta \big(A_{2+}\cos2\psi_o\cos\phi_o - A_{2\times}\sin2\psi_o\sin\phi_o\big),
\\[1ex]
A_{22} =  h_0 \sin^2\theta \big(A_{2+}\sin2\psi_o\cos\phi_o + A_{2\times}\cos2\psi_o\sin\phi_o\big),
\\[1ex]
A_{23} = -h_0 \sin^2\theta \big(A_{2+}\cos2\psi_o\sin\phi_o - A_{2\times}\sin2\psi_o\cos\phi_o\big),
\\[1ex]
A_{24} = -h_0 \sin^2\theta \big(A_{2+}\sin2\psi_o\sin\phi_o + A_{2\times}\cos2\psi_o\cos\phi_o\big).
\end{array}
\ee
In this case the signal is mathematically equivalent to the GW signal from
a biaxial, precessing pulsar, where the angle $\theta$ is the so-called
"wobble angle" (\cite{ZimmermannS1979}) or  from of a spinning, fluid biaxial star that
is not rotating about its principal axis (\cite{BonazzolaG1996}). When we set $\theta = \pi/2$ we obtain
the signal from a triaxial star rotating about one of the principal axes of its moment of
inertia with GW frequency equal to twice the rotational frequency.

\section{$\F$-statistic and the maximum likelihood estimators of the parameters}
\label{sect:stat}
Let us assume that the noise $n(t)$, $t = 1,...,n$ is Gaussian and uncorrelated with the
same variance $\sigma^2$ for each sample $t$ and mean $\mu = 0$.
Let us assume that the signal $s(t)$ present in the data $x(t)$, $t = 1,...,n$ is additive i.e.,
\be
x(t) = n(t) + s(t).
\ee
Let us assume that the signal $s(t)$, $t = 1,...,n$ can be expressed
as a linear combination of known functions $\textsl{h}_l, l = 1,...,L$:
\be
\label{eq:sigl}
s(t) = \sum^L_{l=1} {\mathcal A}_l \textsl{h}_l(t),
\ee
with unknown amplitude parameters ${\mathcal A}_l$. Moreover, let us assume that the variance
$\sigma^2$ of the noise is also unknown. In this case the probability density distribution
(pdf) $p(x)$ of the data is given by
\be
\label{eq:pdf}
p(x;\mathbf{\A},\sigma^2) = \left(\frac{1}{\sqrt{2\pi\sigma^2}}\right)^n\exp\left(-\frac{\sum^n_{t=1}(x(t) - \sum^L_{l=1}\A_l \h_l(t))^2}{2\sigma^2}\right),
\ee
where $\mathbf{\A} = (\A_1,\ldots,\A_L)$.
The likelihood function $\Lambda$ is defined by
\be
\label{eq:Lam}
\Lambda(\mathbf{\A},\sigma^2;x) = p(x;\mathbf{\A},\sigma^2),
\ee
i.e., the likelihood function is just the pdf treated as a function of the parameters of the pdf.
The maximum likelihood estimators  of amplitudes $\A_l$ and of the variance $\sigma^2$
are the values of the parameters that maximize $\Lambda$ and they
are obtained by solving the following set of equations:
\be
\label{eq:mle}
\frac{\partial \Lambda}{\partial \A_{l}} = 0 ,\,\,\, \mbox{for} \,\,\, l=1,...,L,\quad {\rm and}\quad
\frac{\partial \Lambda}{\partial \sigma^2} = 0.
\ee
From Eqs.\,(\ref{eq:pdf}) and (\ref{eq:Lam}) we have
\be
\begin{aligned}
\label{eq:mlee}
\frac{\partial \Lambda}{\partial \A_{l}} &= \Lambda(x) \frac{\sum_{t=1}^n x(t) \h_l(t)
- \sum^L_{l'=1} \A_{l'} \sum^n_{t=1}\sum^n_{t'=1}  \h_{l'}(t') \h_l(t)}{\sigma^2},\\
\frac{\partial \Lambda}{\partial \sigma^2} &= \Lambda(x) \left(-\frac{n}{2}\frac{1}{\sigma^2}
+ \frac{\sum^n_{t=1}(x(t) - \sum^L_{l=1}\A_l \h_l(t))^2}{2(\sigma^2)^2}\right).
\end{aligned}
\ee
From the above equations the maximum likelihood estimators
$\hat{\A}_l$ and $\widehat{\sigma^2}$ read
\bea
\label{eq:mlesol}
\hat{\A}_l &=& \sum_{l'=1}^L M^{-1}_{l'l} N_{l'}, \\
\label{eq:mlesig}
\widehat{\sigma^2} &=& \frac{ \av{x^2} - \sum_{l=1}^L \sum_{l'=1}^L N_l M^{-1}_{l'l} N_{l'} }{n},
\eea
where the operator $\av{\cdot}$ is defined as
\be
\label{eq:tav}
\av{x \ y} := \sum^{n}_{t=1} x(t) y(t)
\ee
and we have introduced a vector $N$ and a matrix $M$ with components
\be
N_l = \av{x \h_l},\quad{\rm and}\quad M_{ll'} = \av{\h_l \h_{l'}}.
\ee
The amplitude estimators $\hat{\A}_l$ are unbiased i.e., $E[\hat{\A}_l] = \A_l$. The maximum likelihood estimator
of variance is biased however, and we have
\be
\label{eq:mlesigbias}
E[\widehat{\sigma^2}] = \frac{n-L}{n}\sigma^2.
\ee

A convenient method for testing the hypothesis of the presence of a signal with unknown
parameters is the {\em likelihood ratio}, or LR test. The likelihood ratio statistic
is given by
\be
\label{LRtest}
LR = 2\log \left[ \frac{\Lambda(\hat\theta;x)}{\Lambda(\hat\theta_r;x)} \right].
\ee
$\Lambda(\hat\theta_r;x)$ is the likelihood function where $r$
parameters $\theta_r$ out of all the unknown parameters $\theta$
are assigned a fixed value.

Asymptotically (i.e., for sample size $n$ approaching to $\infty$), from
Wilks' theorem \cite{Wilks38} the likelihood ratio statistic
is $\chi^2$-distributed with $r$ degrees of freedom. In our case the LR test takes the form
\be
LR = 2\log \left[ \frac{\Lambda(\hat{\A}_l,\widehat{\sigma^2};x)}{\Lambda(\A_l = 0,\sigma^2 = \widehat{\sigma^2}(\A_l=0);x)} \right],
\ee
where we assign fixed values equal to $0$ to the $L$ amplitude parameters $\A_l$.
Using the expression for maximum likelihood estimators obtained above (Eqs.\,(\ref{eq:mlesol})
and (\ref{eq:mlesig}) ) we explicitly have
\be
LR = - n \log\left(1 - 2\frac{\F_{\sigma}}{n}\right),
\ee
where
\bea
\label{eq:Fg}
\F_{\sigma}      &=& \frac{1}{2} \frac{\sum_{l=1}^L \sum_{l'=1}^L N_l M^{-1}_{l'l} N_{l'}}{\sigma^2_r},\\
\label{eq:rawsig}
\sigma^2_r &=& \frac{\av{x^2}}{n}.
\eea
Thus comparing the $LR$ statistic to a threshold is equivalent to comparing statistic
$\F_{\sigma}$ to threshold. The statistic $\F_{\sigma}$ generalizes the well known $\F$-statistic (\cite{JKS98})
to the case when variance of the noise is unknown. The quantity $\sigma^2_r$ is the "raw" estimator
of the variance assuming the data is noise of unknown variance and the known mean equal to 0.
In the case of known variance the likelihood ratio test (Eq.\,\ref{LRtest}) takes the form
\be
LR = 2 \F,
\ee
where $\F$ is the standard $\F$-statistic given by
\be
\F = \frac{1}{2} \frac{\sum_{l=1}^L \sum_{l'=1}^L N_l M^{-1}_{l'l} N_{l'}}{\sigma^2}.
\ee

Let us consider the explicit case of the two component signal given by Eqs.~(\ref{eq:sig1}), (\ref{eq:sig2}) and (\ref{eq:amps}).
The amplitudes $A_{1k}$ and functions $h_{1k}$, ($ k = 1,...,4$) describe
the component of the signal with the GW frequency $\Omega_{GW}$ equal to once the spin frequency,
$\Omega_{GW} = \Omega$ and the amplitudes $A_{2k}$ and functions $h_{2k}$ ($ k = 1,...,4$) describe
the signal with GW frequency equal to twice the spin frequency,
$\Omega_{GW} = 2\Omega$.
Let us assume that we pass the data $x(t)$ through two narrowband filters around the frequencies
$\Omega$ and  $2\Omega$ and as a result we obtain two narrowband data streams $x_1(t)$ and $x_2(t)$,
$t = 1,...,n$. We assume that both data streams are Gaussian, uncorrelated with constant variances $\sigma_1^2$ and $\sigma_2^2$
respectively that are not necessarily equal.
We can assume that the noise samples in the two narrowband data streams are independent and then the probability density
function of data is product of probability functions $p_1$ and $p_2$ for data $x_1$ and $x_2$ respectively.
Thus the likelihood function in the case of a two component signal is given by
\be
\label{eq:Lam2}
\Lambda(\mathbf{\A},\sigma_1^2,\sigma_2^2;x_1,x_2) = p_1(x_1;A_{11},...,A_{14},\sigma_1^2) p_2(x_2;A_{21},...,A_{24},\sigma_2^2).
\ee
By the same derivation as above we obtain that the LR test for the two component signal is equivalent to comparing
to a threshold a statistic  $\F_{\sigma}$ which is the sum of the $\F$-statistics for each component:
\bea
\label{eq:Fgl}
\F_{\sigma}      &=& \frac{1}{2} \frac{\sum_{k=1}^4 \sum_{k'=1}^4 N_{1k} M^{-1}_{1k'k} N_{1k'}}{\sigma^2_{1r}} +
                     \frac{1}{2} \frac{\sum_{k=1}^4 \sum_{k'=1}^4 N_{2k} M^{-1}_{2k'k} N_{2k'}}{\sigma^2_{2r}} ,\\
\label{eq:rawsigl}
\sigma^2_{1r} &=& \frac{\av{x_1^2}}{n}, \,\,\,\,\,\,\,\, \sigma^2_{2r} = \frac{\av{x_2^2}}{n},
\eea
where
\bea
N_{1k} &=& \av{x_1 \h_{1k}},\quad{\rm and}\quad M_{1kk'} = \av{\h_{1k} \h_{1k'}}, \\
N_{2k} &=& \av{x_2 \h_{2k}},\quad{\rm and}\quad M_{2kk'} = \av{\h_{2k} \h_{2k'}}.
\eea
Observing that the amplitude modulation functions $a$ and $b$
vary much more slowly than the phase $\phi$ of the signal
and assuming that the observation time is much longer than
the period of the signal we approximately have (see \cite{JKS98} for details)
\be
\label{app1}
\begin{array}{l}
\av{h_{l1}\,h_{l3}} \cong \av{h_{l1}\,h_{l4}} \cong \av{h_{l2}\,h_{l3}} \cong \av{h_{l2}\,h_{l4}} \cong 0,
\\[1ex]
\av{h_{l1}\,h_{l1}} \cong \av{h_{l3}\,h_{l3}} \cong \frac{1}{2} A, \quad
\av{h_{l2}\,h_{l2}} \cong \av{h_{l4}\,h_{l4}} \cong \frac{1}{2} B, \quad
\av{h_{l1}\,h_{l2}} \cong \av{h_{l3}\,h_{l4}} \cong \frac{1}{2} C,
\end{array}
\ee
(for $l = 1,2$), where we have introduced the time averages
\be
\label{ABCdef}
A := \av{a^2}, \quad
B := \av{b^2}, \quad
C := \av{ab}.
\ee

With the above approximations the maximum likelihood estimators of
the amplitudes $A_{lk}, (l=1,2, k=1,\ldots,4)$ given by
Eq.\,(\ref{eq:mlesol}) take the following explicit form:
\be
\begin{aligned}
\label{eq:ampmle}
\hat{A}_{11} \cong &\,\frac{2}{D} ( B \av{x h_{11}} - C \av{x h_{11}}),
\quad
\hat{A}_{12} \cong \,\frac{2}{D} ( A \av{x h_{12}} - C \av{x h_{11}}),\\
\hat{A}_{13} \cong &\,\frac{2}{D} ( B \av{x h_{13}} - C \av{x h_{13}}),
\quad
\hat{A}_{14} \cong \,\frac{2}{D} ( A \av{x h_{14}} - C \av{x h_{14}}),\\
\hat{A}_{21} \cong &\,\frac{2}{D} ( B \av{x h_{21}} - C \av{x h_{21}}),
\quad
\hat{A}_{22} \cong \,\frac{2}{D} ( A \av{x h_{22}} - C \av{x h_{21}}),\\
\hat{A}_{23} \cong &\,\frac{2}{D} ( B \av{x h_{23}} - C \av{x h_{23}}),
\quad
\hat{A}_{24} \cong \,\frac{2}{D} ( A \av{x h_{24}} - C \av{x h_{24}}).
\end{aligned}
\ee
where $D:=AB-C^2$. Using the approximation given by Eqs.\,(\ref{app1}), the generalized $\F$-statistic $\F_{\sigma}$
splits into the sum of two $\F$-statistics corresponding to the two components of the signal:
\be
\label{eq:Fstatg}
\F_{\sigma} =  \F_{1\sigma} + \F_{2\sigma},
\ee
where
\be
\label{eq:Fstat}
\F_{l\sigma} \cong  \frac{
\Big( B\, (\av{x h_{l1}}^2 + \av{x h_{l3}}^2) + A\, (\av{x h_{l2}}^2 + \av{x h_{l4}}^2)
- 2C\, (\av{x h_{l1}} \av{x h_{l2}} + \av{x h_{l3}} \av{x h_{l4}}) \Big)}{D \widehat{\sigma_l^2}_r}
\ee
for $l = 1,2$, with $\widehat{\sigma_l^2}_r$ given by Eq.\,(\ref{eq:rawsigl}).
\section{The signal-to-noise ratio}
\label{sect:snr}
For a signal $s$ added to a Gaussian noise with variance $\sigma^2$,
the signal-to-noise ratio is given by
\be
\rho^2 = \frac{ \av{ s^2 } }{\sigma^2}.
\ee
In the case of a signal consisting of two components (Eqs.~\ref{eq:sig1} and \ref{eq:sig2}) 
one has, assuming that the cross-correlation terms between the two
components vanish, 
\be
\rho^2 = \rho^2_1 + \rho^2_2,
\ee
where $\rho_1$ and $\rho_2$ are the signal-to-noise ratios of the two components:
\be
\rho_1^2 = \frac{ \av{ s_1^2 }}{\sigma_1^2}, \hspace{3mm}  \rho_2^2 = \frac{ \av{ s_2^2 }}{\sigma_2^2}.
\ee

\begin{figure}
\begin{center}
\scalebox{0.65}{\includegraphics{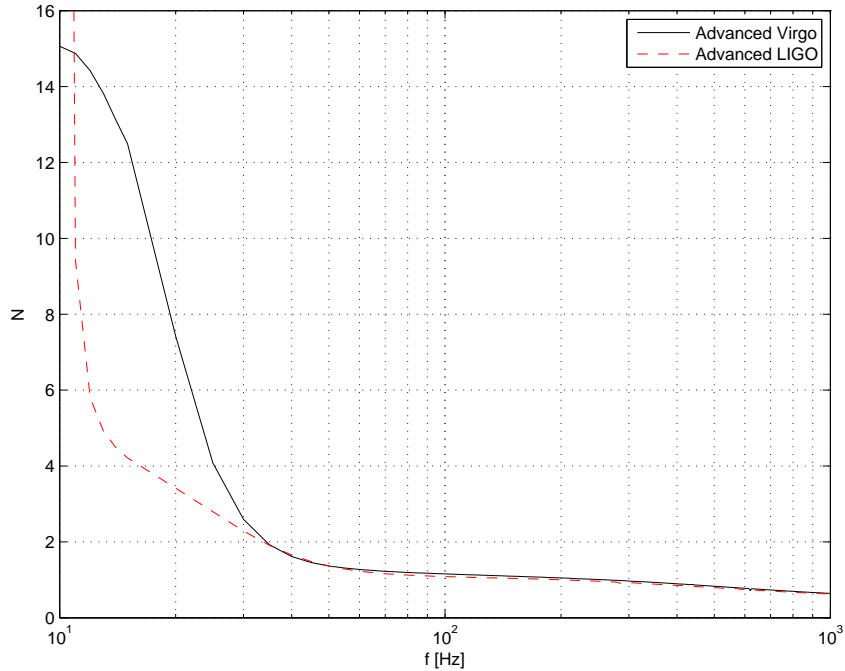}}
\caption{\label{snr_adv}
Ratio $\mathbf{N}$ of the amplitude spectral densities at frequency $f$ and at frequency $2f$ for
Advanced Virgo (continuous line) and Advanced LIGO (dashed line) detectors as a function of frequency $f$.}
\end{center}
\end{figure}

Signal-to-noise ratio for our signal is independent of the phase angle $\phi_o$. It is however a very complicated function of the angles $\alpha,\delta,\psi_o,\iota,\psi$ and $\theta$, hence it is useful to obtain quantities averaged over the angles. Averaging is performed according to the following definition:
\be
\label{av}
\langle\cdots\rangle_{\alpha,\delta,\psi_o,\iota,\psi,\theta} :=
\frac{1}{2\pi}\int_0^{2\pi}d\alpha
\times \frac{1}{2}\int_{-1}^{1}d\sin\delta
\times\frac{1}{2\pi}\int_0^{2\pi}d\psi_o
\times\frac{1}{2}\int_{-1}^{1}d\cos \iota
\times\frac{1}{2\pi}\int_0^{2\pi}d\psi
\times\frac{1}{\pi}\int_{0}^{\pi}d\theta \left(\cdots\right).
\ee
Note that because $\delta\in[-\pi/2,\pi/2]$, the integration over $\sin\delta$
rather than $\cos\delta$ is performed in Eq.\ (\ref{av}).  Let us consider the
averages about the sky position of the source given by the angles $\alpha$ and
$\delta$, and the polarization given by the angles $\psi_o$ and $\iota$.  We
find that these averages are independent of the position of the detector on
Earth and the orientation of its arms:
\bea
\label{eq:snr1}
\langle\rho_1^2\rangle_{\alpha,\delta,\psi_o,\iota,\psi,\theta} &=& \frac{1}{400} (h_0^2 + \frac{7}{8} h_1^2  - h_1 h_0)\frac{n}{\sigma_1^2}, \\
\label{eq:snr2}
\langle\rho_2^2\rangle_{\alpha,\delta,\psi_o,\iota,\psi,\theta} &=& \frac{3}{100} (h_0^2 + \frac{41}{24} h_1^2  - h_1 h_0)\frac{n}{\sigma_2^2},
\eea
where $n$ is the number of data points.
It is useful to see what is the ratio $\mathcal{S} = \sqrt{\langle\rho_1^2\rangle_{\alpha,\delta,\psi_o,\iota,\psi,\theta}/\langle\rho_2^2\rangle_{\alpha,\delta,\psi_o,\iota,\psi,\theta}}$
of the average SNRs of the two components. It can be expressed as a function of the ratio $R$ of the two amplitudes,
$R = h_0/h_1 = (I_3 - I_1)/(I_2 - I_1)$ and the ratio $\mathbf{N}$ of the variances of noise around the two components,
$\mathbf{N} = \sigma_1/\sigma_2$: 
\be
\label{eq:S}
\mathcal{S} = \mathcal{R}/\mathbf{N},
\ee
where
\be
\mathcal{R} = \sqrt{\frac{1}{12} \frac{R^2 - R + \frac{7}{8}}{R^2 - R + \frac{41}{24}}}.
\ee
Let us first consider the case when the variances of noise for the two components are equal, i.e., ${\bf N} = 1$.
Then the SNR is determined by factor $\mathcal{R}$.
We find that the average SNR (Eq.\,\ref{eq:snr1}) of the $\Omega$ component is always less than that
of the $2 \Omega$ component (Eq.\,\ref{eq:snr2}). We find that $\mathcal{S}$ reaches the maximum
of around $0.28$ when amplitude $h_1 = 0$ ($R = \infty$) and it has minimum of around 0.19 when $h_0 = h_1/2$
($R=1/2$). When the amplitude $h_0$ vanishes (i.e., for $R = 0$), $\mathcal{S} \simeq 0.21$.
It is useful to consider the ratio of SNRs of the two components taking into account planned Advanced detectors
sensitivity curves. For a given frequency $f$ the ratio $\mathbf{N}$ is equal to the ratio of amplitude spectral
densities at frequencies $f$ and $2 f$.
In Figure \ref{snr_adv} we have plotted the ratio $\mathbf{N}$ as a function of frequency
of $1 \Omega$ component for the advanced Virgo and advanced LIGO detectors. The sensitivity curves considered are the final design sensitivity curves given in Fig. 1  of \cite{LSC13a}.
We have considered the range
of frequencies from 10Hz to 1kHz corresponding to the range of twice the frequency of 20Hz to 2kHz.
The $1\Omega$ component dominates ($\mathbf{N} < 1$, i.e., $\sigma_1 < \sigma_2$) 
for frequencies greater than 200Hz and 260Hz for Advanced LIGO
and Advanced Virgo detectors, respectively. For frequencies below 30Hz 
the $1\Omega$ component is very much suppressed as compared to the $2\Omega$ one.

For the special case when the two moments of inertia $I_1$ and $I_2$ are equal, the averages over the position angles and polarization angles read
\bea
\langle\rho_1^2\rangle_{\alpha,\delta,\psi_o,\iota} &=&  \frac{1}{200} h_0^2\sin^2 2\theta\frac{n}{\sigma^2},\\
\langle\rho_2^2\rangle_{\alpha,\delta,\psi_o,\iota} &=&  \frac{2}{25}  h_0^2\sin^4  \theta\frac{n}{\sigma^2},
\eea
and they are equivalent to Eqs.\,(94) and (95) of \cite{JKS98}, assuming that the following relations hold:
$S_h(f_o) = 2\sigma_1^2 \Delta t$ $S_h(2 f_o) = 2\sigma_2^2 \Delta t$,
where $S_h(f)$ is one-sided spectral density at frequency $f$ and $\Delta t$ is the sampling time.

Assuming that spectral densities at two frequencies are equal, 
for small values of the angle $\theta$ the average SNR of the $1 \Omega$ component is greater than the one for $2 \Omega$ component. The two SNRs become equal for $\theta_0 \simeq 27\deg$, and for $\theta > \theta_0$ the $2 \Omega$ component dominates.

\section{The Fisher matrix}
\label{sect:fisher}
In the analysis of the estimation method proposed below we shall use the Fisher matrix
to assess the accuracy of the parameter estimators.
We have two theorems (see e.g., \cite{LC98}, Theorem 6.6 p. 127, and Theorem 5.1 p. 463) that can loosely be stated as follows:

\newtheorem{theorem}{Theorem}

\begin{theorem}[Cram\`er-Rao bound]
The diagonal elements of the inverse of the Fisher matrix
are lower bounds on the variances of unbiased estimators of the parameters.
\end{theorem}

\begin{theorem}
Asymptotically (i.e., when signal-to-noise ratio tends to infinity)
the ML estimators are unbiased, normally distributed and their covariance matrix is equal
to the inverse of the Fisher matrix.
\end{theorem}

For a signal $s=s(t;\btheta)$ added to a Gaussian noise with variance $\sigma^2$
which depends on $M$ parameters $\btheta=(\theta_1,\ldots,\btheta_M)$,
the elements of the Fisher matrix $\Gamma(\btheta)$ are given by
\be
\label{mFisher}
\Gamma_{{\theta_i}{\theta_j}} = \frac{1}{\sigma^2}
\av{\frac{\partial s}{\partial\theta_i}\frac{\partial s}{\partial\theta_j}},
\quad i,j=1,\ldots,M.
\ee

For a signal $s(t)$ (see Eq.\,\ref{eq:sigl}) which is a linear
function of $L$ amplitudes $\A_l$ and the amplitudes depend on $M$ parameters $\theta_m$ 
($m=1,\ldots,M$) 
it is convenient to calculate the Fisher matrix $\Gamma(\btheta)$ using the following formula
($\mathsf{T}$ denotes here matrix transposition):
\be
\Gamma(\btheta) = J^\mathsf{T} \cdot \Gamma(\mathbf{\A})\cdot J,
\ee
where the Jacobi $L \times M$ matrix $J$ has elements $\partial \A_l/\partial\theta_m$
($l=1,\ldots,L$, $m=1,\ldots,M$).
The components of the matrix $\Gamma(\mathbf{\A})$ are given by
\be
\label{AFisher}
\Gamma_{{\A_l}{\A_{l'}}} = \frac{1}{\sigma^2}
\av{\frac{\partial s}{\partial \A_l}\frac{\partial s}{\partial \A_{l'}}},
\quad l,l'=1,\ldots,L.
\ee

For our two component signal model given by Eqs.~(\ref{eq:sig1}), (\ref{eq:sig2}) and (\ref{eq:amps})
and with the approximations given by Eqs.\,\eqref{app1}, the matrix $\Gamma(\mathbf{\A})$  is given by
\be
\label{eq:Mmat}
\Gamma(\mathbf{\A}) =
\begin{pmatrix}
{\cal M} & 0 \\[1ex]
0 & {\cal M}
\end{pmatrix}
\quad
{\rm where}
\quad
{\cal M} = \frac{1}{2\sigma^2}
\begin{pmatrix}
A & C & 0 & 0 \\[1ex]
C & B & 0 & 0 \\[1ex]
0 & 0 & A & C \\[1ex]
0 & 0 & C & B
\end{pmatrix},
\ee
and where $A, B, C$ are defined by Eqs.\,(\ref{ABCdef}).
Thus the $8 \times 8$ matrix $\Gamma(\mathbf{\A})$ splits into two identical components defined
by the $4 \times 4$ matrix  ${\cal M}$.


\section{Data analysis method}
\label{sect:DataMethod}
Let us consider the  signal $s(t)$ given by  Eq.\,(\ref{eq:sigl})
which is a linear function of $L$ amplitudes $\A_l$
and let us assume that the amplitudes depend on $M$ independent parameters $\theta_m$ ($L > M$).
To detect the signal we use the likelihood ratio test which is equivalent to comparing
the $\F_{\sigma}$-statistic given by Eqs.\,(\ref{eq:Fg}) to a threshold.
If the value of $\F_{\sigma}$ is statistically significant we may estimate the parameters.
First we obtain the maximum likelihood estimators of the amplitude parameters $\A_l$ using the explicit
analytic formula  given by Eqs. (\ref{eq:mlesol}).
Then we obtain estimators of the independent parameters by a least squares
fit i.e., estimators of the parameters are obtained by minimizing the following function $LS$ with respect to M
parameters $\theta_m$:
\be
LS = \sum_{l=1}^L\sum_{l'=1}^L [ \hat{\A_{l}} - \A_{l}(\theta_1,\ldots,\theta_M) ]\, \Gamma_{{\A_l}{\A_{l'}}} \,
                               [ \hat{\A_{l'}}  - \A_{l'}(\theta_1,\ldots,\theta_M) ].
\ee
For our two component signal model given by Eqs.~(\ref{eq:sig1}), (\ref{eq:sig2}) and (\ref{eq:amps})
the function $LS$ becomes
\bea
\label{eq:LS}
LS &=& \sum_{k=1}^4\sum_{k'=1}^4 ( \hat{A}_{1k} - A_{1k}(h_0,h_1,\phi_o,\psi_o,\iota,\theta,\psi) )\, {\cal M}_{kk'} \,
                                 ( \hat{A}_{1k'} -A_{1k'}(h_0,h_1,\phi_o,\psi_o,\iota,\theta,\psi) ) \nonumber \\
   &+& \sum_{k=1}^4\sum_{k'=1}^4 ( \hat{A}_{2k} - A_{2k}(h_0,h_1,\phi_o,\psi_o,\iota,\theta,\psi) )\, {\cal M}_{kk'} \,
                                 ( \hat{A}_{2k'} -A_{2k'}(h_0,h_1,\phi_o,\psi_o,\iota,\theta,\psi) ),
\eea
where ${\cal M}_{kk'}$ are components of the $4 \times 4$  matrix $\cal{M}$ given by Eq.\,(\ref{eq:Mmat}).
The least squares fit involves a non-linear minimization procedure for which we need the initial values
for the 6 parameters $(\iota,\psi_o,G_1,G_2,H_1,H_2)$ with respect to which the $LS$ function is minimized.
For the initial values we use an analytic solution for the six parameters
in terms of the amplitude parameters $A_{lk}$, $(l=1,2, k=1,\ldots,4)$.
Many such solutions exist. We use the following;
to present it in a compact form we first introduce the auxiliary quantities
\bea
E_1 &=& A_{11}^2 + A_{12}^2 + A_{13}^2 + A_{14}^2,\\
I_1 &=& A_{11} A_{14} - A_{12} A_{13},\\
h_{1c} &=& \sqrt{(E_1 + \sqrt{E_1^2 - 4 I_1^2})/2},\\
h_{1p} &=&  \mbox{sign}(I_1) \sqrt{(E_1 - \sqrt{E_1^2 - 4 I_1^2})/2},\\
E_2 &=& A_{21}^2 + A_{22}^2 + A_{23}^2 + A_{24}^2,\\
I_2 &=& A_{21} A_{24} - A_{22} A_{23},\\
h_{2c} &=& \sqrt{(E_2 + \sqrt{E_2^2 - 4 I_2^2})/2},\\
h_{2p} &=&  \mbox{sign}(I_2) \sqrt{(E_2 - \sqrt{E_2^2 - 4 I_2^2})/2}.
\eea
Let us also introduce the following quantities that depend on
polarization angles $\iota$ through quantities defined by Eqs.~(\ref{aa}):
\be
\begin{aligned}
&b_1 = - A_{1\times} A_{11} + A_{1+} A_{14} , \qquad
b_2 = A_{1\times} A_{12} + A_{1+} A_{13}, \qquad
b_3 = A_{1\times} A_{14} - A_{1+} A_{11}.
\end{aligned}
\ee
\be
\begin{aligned}
&c_1 = - A_{2\times} A_{21} + A_{2+} A_{24} , \qquad
c_2 = A_{2\times} A_{22} + A_{2+} A_{23}, \qquad
c_3 = A_{2\times} A_{24} - A_{2+} A_{21}.
\end{aligned}
\ee
For the polarization angles $\iota$ and $\psi_o$ we obtain:
\be
\label{eq:polang}
\cos\iota = \frac{ h_{2p} }{ h_{2c} + \sqrt{h_{2c}^2 - h_{2p}^2} }, \quad
\psi_o = \frac{1}{2}\arctan\frac{c_1}{c_2}.
\ee
Analytic expression for parameters $G_1, G_2, H_1, H_2$ in terms of the amplitude parameters and
polarization angles obtained are given by
\be
\label{eq:G_an}
G_1  = \frac{4 b_3}{\sin^4\iota \cos 2\psi_o},\qquad
G_2  = \frac{4 b_2}{\sin^4\iota \cos 2\psi_o},
\ee
\be
\label{eq:H_an}
H_1 = -\frac{4 c_3}{\sin^4\iota \cos 2\psi_o},\qquad
H_2 = -\frac{4 c_2}{\sin^4\iota \cos 2\psi_o},
\ee
where we assume that the denominator $\sin^4\iota \cos 2\psi_o$ is not equal to 0.
In the case of the five parameter model described by Eqs.\,(\ref{eq:ampone}) an analytic solution for polarization angles $\iota,\psi_o$
is also given by Eqs.~(\ref{eq:polang}) above.
The solution for the phase angle $\phi_o$ reads
\be
\label{eq:phi5}
\phi_o = \frac{1}{2}\arctan\frac{c_2}{c_3},
\ee
The angle $\theta$ and the amplitude $h_0$ are given by
\be
\label{eq:h0theta5}
h_0 = \frac{g_1^2 + g_2^2}{g_2} ,\qquad
\theta = \arctan(\sqrt{g_2/g_1}),
\ee
where
\bea
g_1 = \frac{2 h_{1c}^2}{\sqrt{h_{1c}^2 - h_{1p}^2}},\\
g_2 = h_{2c} + \sqrt{h_{2c}^2 - h_{2p}^2}.
\eea
In the case of the 4 parameter model for a GW signal from a triaxial star rotating about its principal axis
there is a unique solution for the $(h_0,\phi_o,\psi_o,\iota)$ parameters in terms of the 4 amplitudes
$A_{2k}, k = 1,2,3,4$ and it is given by
\bea
\label{eq:mlea1}
\cos\iota &=& \frac{h_{2p}}{h_{2c} + \sqrt{h_{2c}^2 - h_{2p}^2}}, \\
\label{eq:mlea2}
\psi_o &=& \frac{1}{2}\arctan\frac{c_1}{c_2}, \\
\label{eq:mlea3}
\phi_o &=& \frac{1}{2}\arctan\frac{c_2}{c_3}, \\
\label{eq:mlea4}
h_o &=& h_{2c} + \sqrt{h_{2c}^2 - h_{2p}^2}.
\eea

Using the above Eqs.\,(\ref{eq:mlea1}) - (\ref{eq:mlea1}) one obtains
{\it directly} the maximum likelihood estimators of the four astrophysical
parameters $(h_0,\phi_o,\psi_o,\iota)$  from the maximum likelihood
estimators of the four amplitudes $A_{2k}, k = 1,2,3,4$. 

\section{Monte Carlo simulations}
\label{sect:mcsim}

We have carried out the Monte Carlo simulations in order to test the performance of the estimation method
proposed in Section \ref{sect:DataMethod}.
Each simulation consisted of generating a signal and adding it to white, Gaussian noise and then applying
our algorithm to estimate the parameters of the signal. For the case of a two component model and two narrowband data streams
we assumed for simplicity that the variance of noise for each data stream is the same. We have added signals with signal-to-noise ratios
ranging from  1 to 20. The added signals had both the amplitude and the phase modulation. The phase modulation includes  
the Doppler modulation and two spindowns.
For each signal-to-noise ratio the simulation run was repeated 1000 times for different realizations of the noise.
Then the mean values and variances of the parameter estimators were calculated. We have compared the mean values
with the true values of the injected parameters, as well as compared the variances of the parameters with
the asymptotic values  given by diagonal elements of the inverse of the Fisher matrix for a given signal.
Three signal models were considered: GW signal at twice the spin frequency from a triaxial ellipsoid
spinning about its principal axis, signal with two components at once and twice the spin frequency
from a biaxial star with its spin and principal axes misaligned (Eq.\,\ref{eq:ampone}), and a general two component model of a triaxial star not spinning about its principal axis (Eqs.\,\ref{eq:sig1} - \ref{aa}).

\begin{figure}
\begin{center}
\scalebox{0.65}{\includegraphics{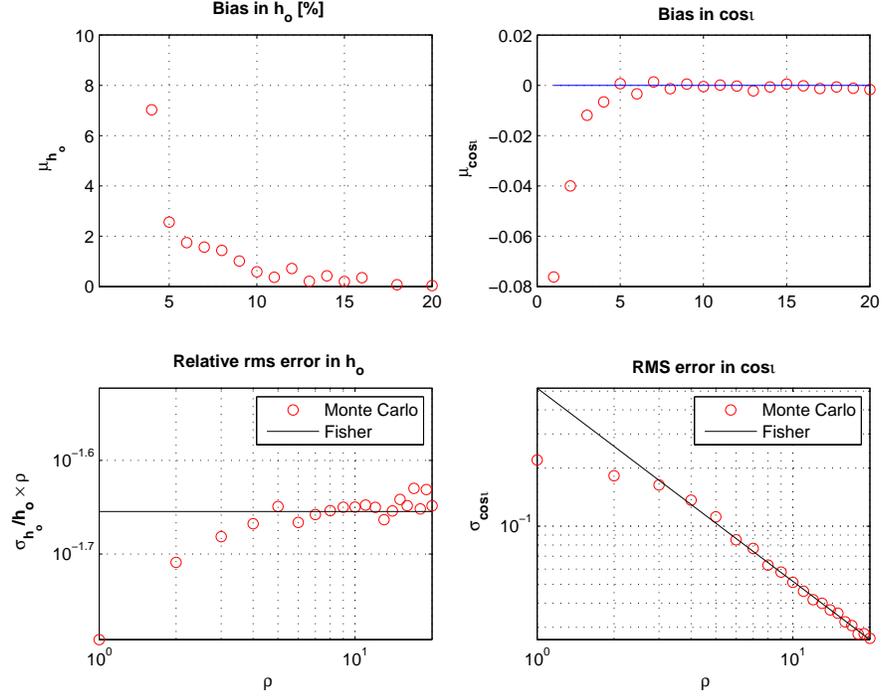}}
\caption{\label{fig:MC2f_fig1}
Four parameter model. The mean and standard deviation of the ML estimator of the amplitude $h_0$
and cosine inclination angle $\cos\iota$  as functions of the SNR.
Top two panels show biases of the estimators, bottom two panels show the standard deviations.
The circles are the results of the simulation whereas continuous lines are
obtained form the Fisher matrix calculations for various SNRs $\rho$. For the case of amplitude
$h_0$ we give a relative error multiplied by the signal-to-noise ratio.}
\end{center}
\end{figure}

\begin{figure}
\begin{center}
\scalebox{0.65}{\includegraphics{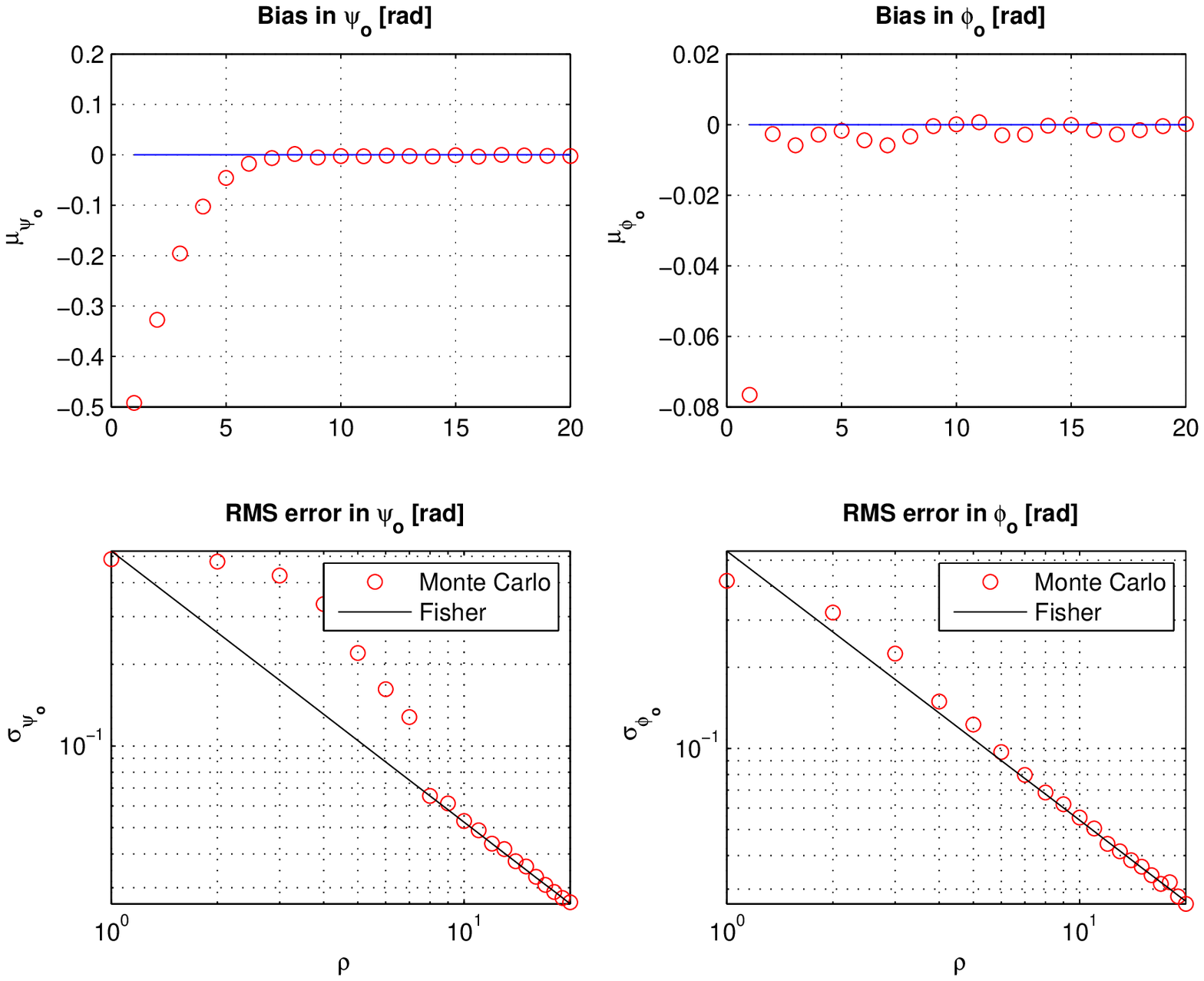}}
\caption{\label{fig:MC2f_fig2}
Four parameter model. The mean and standard deviation of the ML estimator of
the polarization angle $\psi_o$ and the phase angle $\phi_o$ as functions of the SNR.
Top two panels show biases of the estimators, bottom two panels show the standard deviations.}
\end{center}
\end{figure}

\begin{figure}
\begin{center}
\scalebox{0.65}{\includegraphics{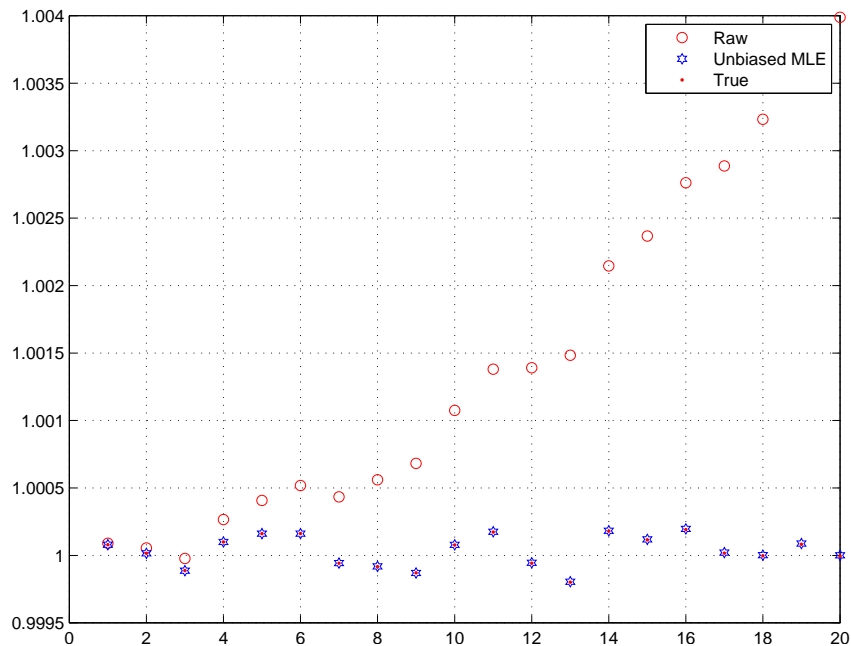}}
\caption{\label{fig:MC2f_fig3}
Four parameter model. Estimation of variance. Blue circles are ML estimators of
the variance of the noise given by Eq.\,(\ref{eq:mlesig}) for various signal-to-noise ratios,
red dots are true variances, and red circles are raw estimates $\sigma^2_r$ of the variance
given by Eq.\,(\ref{eq:rawsig}).}
\end{center}
\end{figure}

The first model has 4 parameters and relevant Monte Carlo simulations are given in Figures \ref{fig:MC2f_fig1}, \ref{fig:MC2f_fig2}, and \ref{fig:MC2f_fig3}.
Here there was no need for the least squares fit, the maximum likelihood estimators of parameters
$(h_0,\phi_o,\psi_o,\iota)$ were calculated from the analytic formulas (\ref{eq:mlea1}) - (\ref{eq:mlea4}).
In Figure \ref{fig:MC2f_fig1} we present biases and standard deviations
as functions of the signal-to-noise ratio of the injected signal
for the amplitude $h_0$ and the inclination angle $\iota$, whereas in Figure \ref{fig:MC2f_fig2} we present the results
for angles $\psi_o$ and $\phi_o$. We find that our estimators, above a signal-to-noise ratio of around 8
are to a very good accuracy unbiased and their variances are very close to the ones calculated from the inverse of the Fisher matrix.
In Figure \ref{fig:MC2f_fig3} we show the variance of the noise estimation results using an unbiased estimator $\widehat{\sigma^2}_u$ obtained form the maximum likelihood (ML) estimator of variance given by Eq.\,(\ref{eq:mlesig}):
\be
\label{eq:sigu}
\widehat{\sigma^2}_u = \frac{n}{n-L}\widehat{\sigma^2}
\ee
For each signal-to-noise ratio we plot the means of the variances of the noise from the
1000 simulations and compare them with the mean values of the unbiased estimators of the variance
given by Eq.\,(\ref{eq:sigu}).

\begin{figure}
\begin{center}
\scalebox{0.65}{\includegraphics{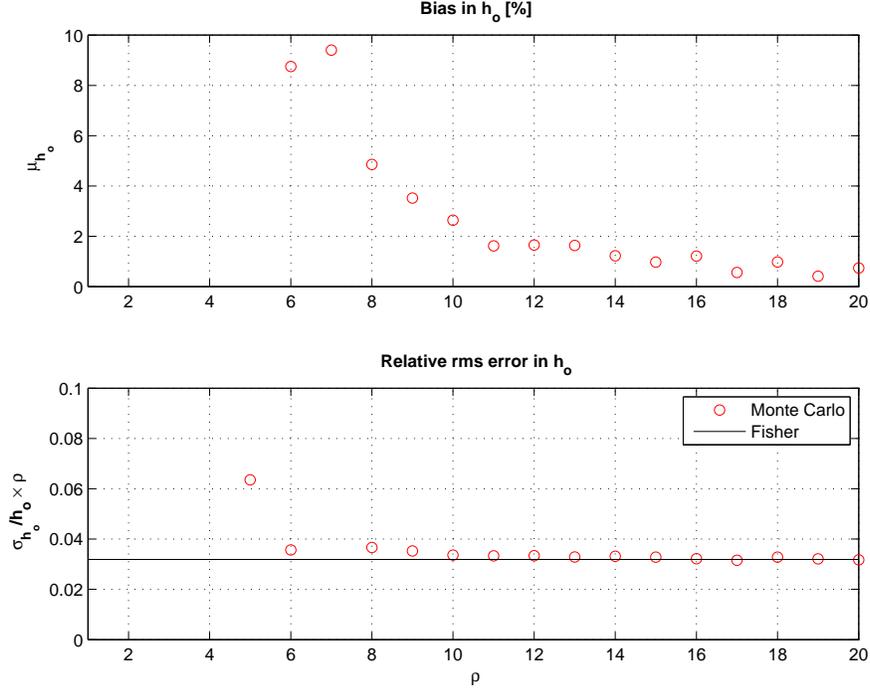}}
\caption{\label{fig:MC1f2f5_fig1}
Five parameter model. The mean and standard deviation of the amplitude $h_0$
estimator as functions of the SNR. The estimator is obtained by the least squares
fit to the ML estimators of the amplitudes $A_l$ as described in Section \ref{sect:DataMethod}.
Top panel shows the biases of the estimators and the bottom panel shows the standard deviation.
Circles are the results of the simulation whereas continuous lines are
obtained form the Fisher matrix calculations for various SNRs $\rho$. For the amplitude
$h_0$ we plot a relative error multiplied by the signal-to-noise ratio.}
\end{center}
\end{figure}

\begin{figure}
\begin{center}
\scalebox{0.65}{\includegraphics{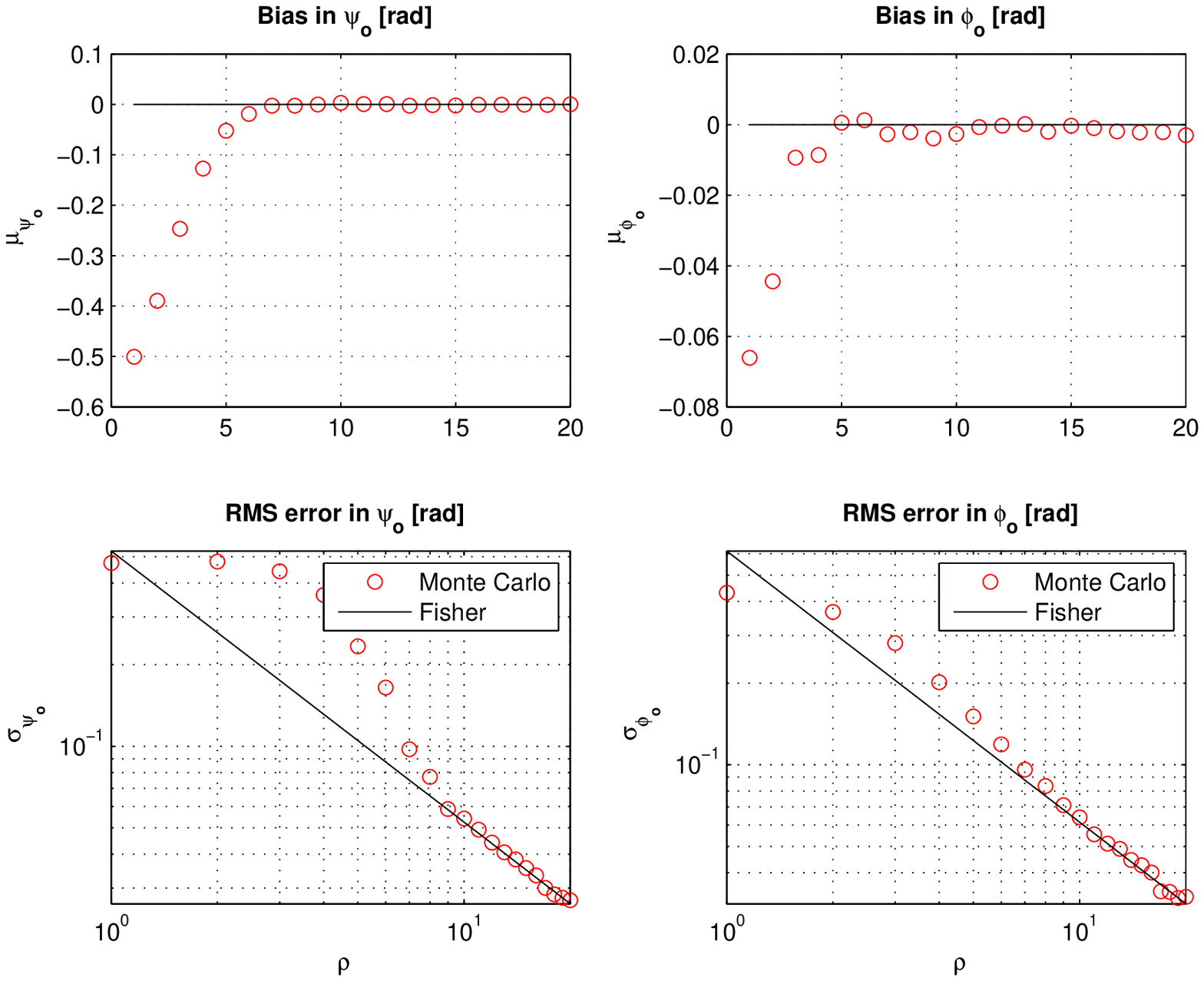}}
\caption{\label{fig:MC1f2f5_fig2}
Five parameter model. The mean and standard deviation of the estimators of the
polarization angle $\psi_o$ and the phase angle $\phi_o$ as functions of the SNR.
Top two panels show biases of the estimators, bottom two panels show the standard deviations.}
\end{center}
\end{figure}

\begin{figure}
\begin{center}
\scalebox{0.65}{\includegraphics{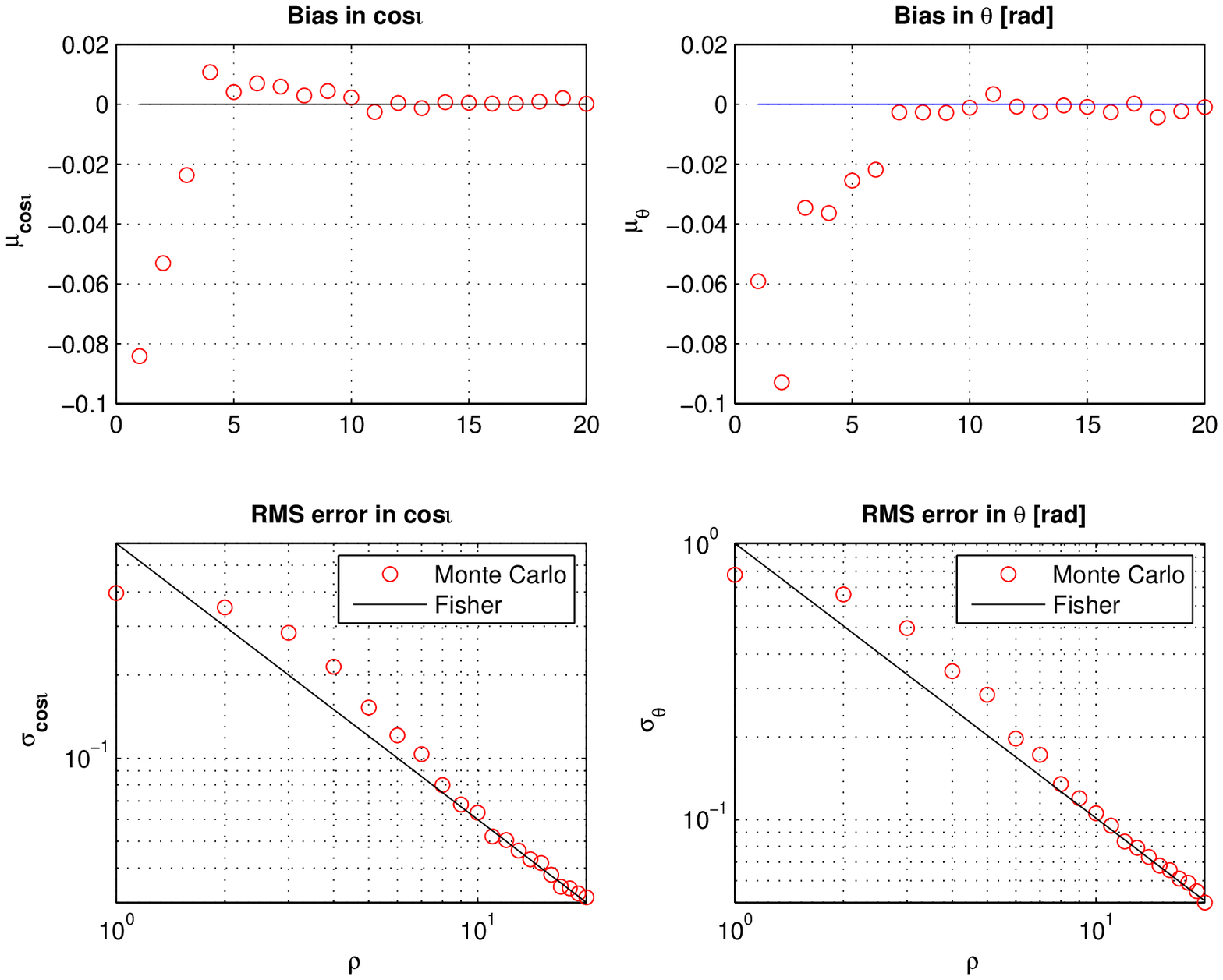}}
\caption{\label{fig:MC1f2f5_fig3}
Five parameter model. The mean and standard deviation of the estimators of the cosine inclination angle
$\cos\iota$ and the wobble angle $\theta$ as functions of the SNR.
Top two panels show biases of the estimators, bottom two panels show the standard deviations.}
\end{center}
\end{figure}

\begin{figure}
\begin{center}
\scalebox{0.65}{\includegraphics{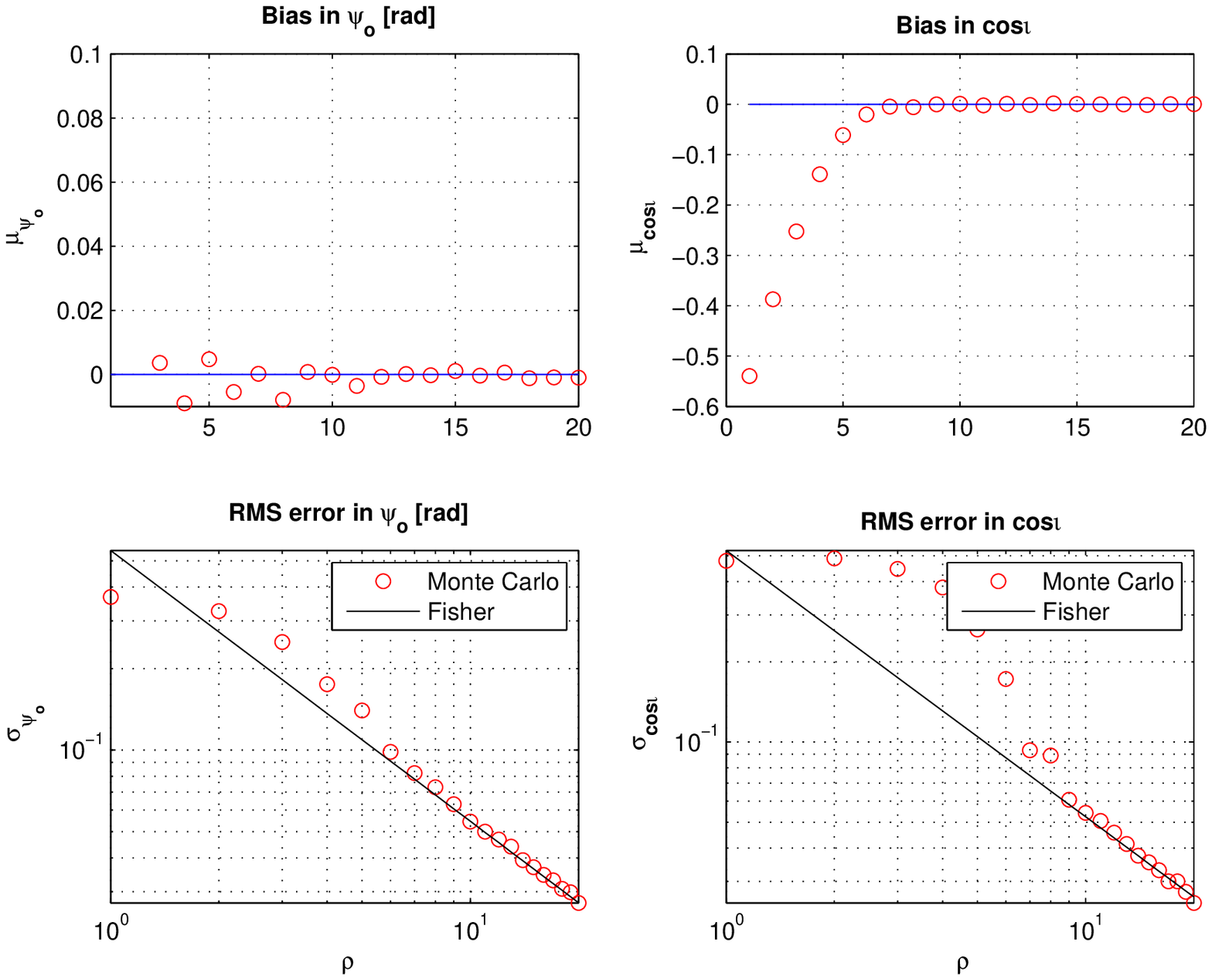}}
\caption{\label{fig:MC1f2f6_fig1}
Six parameter model. The mean and standard deviation of the estimators of the
polarization angle $\psi_o$ and the cosine inclination angle
$\cos\iota$  as functions of the SNR. Top two panels show biases of the estimators,
bottom two panels show the standard deviations.}
\end{center}
\end{figure}

\begin{figure}
\begin{center}
\scalebox{0.65}{\includegraphics{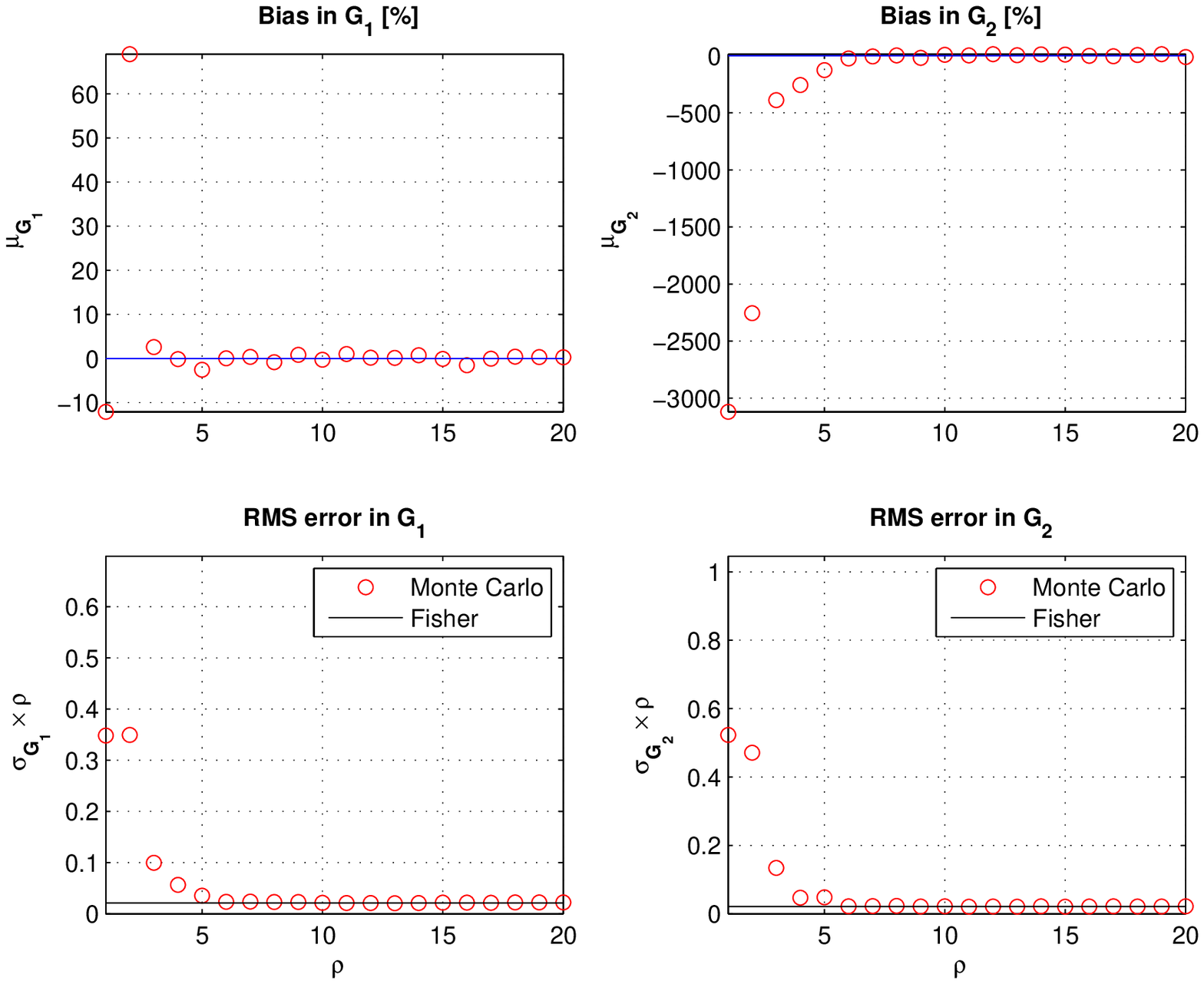}}
\caption{\label{fig:MC1f2f6_fig2}
Six parameter model. The mean and standard deviation of the estimators of the
parameters $G_1$ and $G_2$ (defined by Eqs.\,\ref{eq:G}) as functions of the SNR.
Top two panels show biases of the estimators,
bottom two panels show the standard deviations.}
\end{center}
\end{figure}

\begin{figure}
\begin{center}
\scalebox{0.65}{\includegraphics{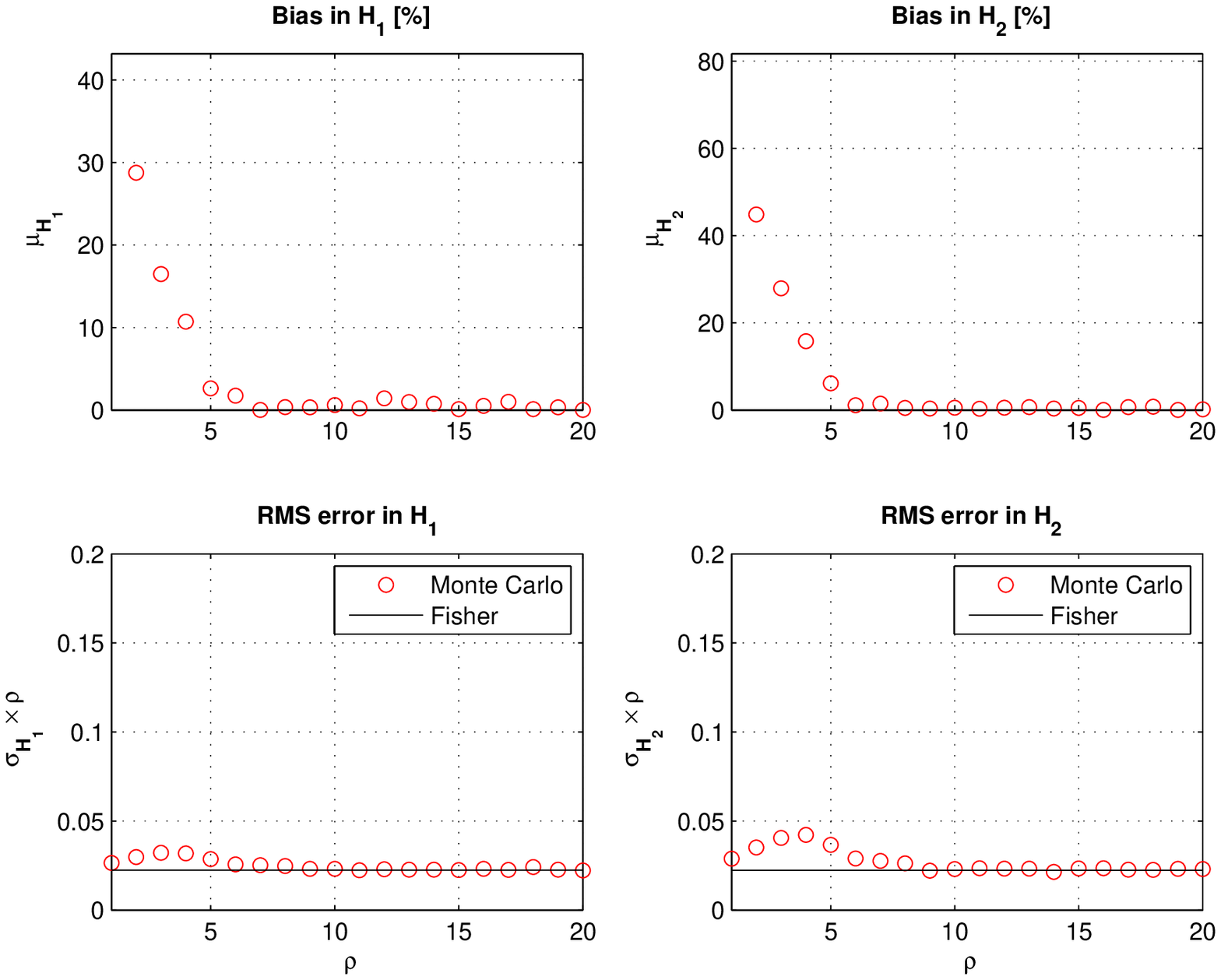}}
\caption{\label{fig:MC1f2f6_fig3}
Six parameter model. The mean and standard deviation of the estimators of the
parameters $H_1$ and $H_2$ (defined by Eqs.\,\ref{eq:H}) as functions of the SNR.
Top two panels show biases of the estimators,
bottom two panels show the standard deviations.}
\end{center}
\end{figure}

In Figures \ref{fig:MC1f2f5_fig1}, \ref{fig:MC1f2f5_fig2} and \ref{fig:MC1f2f5_fig3}
we present the results of a simulation for 5 parameter model with 8 amplitude given
by Eqs.\,(\ref{eq:ampone}). We estimate the five parameters $h_o,\iota,\psi_o,\phi_o,\theta$
by minimizing the function $LS$ (Eq.\,\ref{eq:LS}) with the initial values
for the parameters given by Eqs.\,(\ref{eq:polang}), (\ref{eq:phi5}) and (\ref{eq:h0theta5}).
We find again that above the signal-to-noise ratio
of around 8 our estimators are almost unbiased and their variances are closely reproduced by
the Fisher matrix.
Finally we consider the 6 parameter model for the core superfluid component "pinning" to the solid crust
recently proposed by Jones \cite{Jones2010}, where the 8 amplitude parameters are given by Eqs.\,(\ref{eq:amp8}).
Here to estimate the 6 parameters  $(\iota,\psi_o,G_1,G_2,H_1,H_2)$  we employ our least
squares procedure with the initial values for the parameters given by an analytic solution presented by
Eqs.\,(\ref{eq:polang}), (\ref{eq:G_an}) and (\ref{eq:H_an}). The results of the Monte Carlo simulations are given in Figures \ref{fig:MC1f2f6_fig1}, \ref{fig:MC1f2f6_fig2} and \ref{fig:MC1f2f6_fig3}. Like in the previous 5 parameter case
for SNR above around 8 the estimators are almost unbiased and their variances are close the
variance defined by the diagonal element of the inverse of the Fisher matrix for the signal model.

\section{Conclusions}
\label{sect:concl}
We have proposed a method to estimate the parameters of a GW signal from a known pulsar
assuming that the signal is emitted at both once and twice the spin frequency. Our method involves representing the signal as a linear function of 8 amplitude parameters. The 8 amplitudes are functions
of the astrophysical parameters. The scheme involved obtaining the maximum likelihood estimators
of the 8 amplitude parameters first, and then obtaining the estimators of the astrophysical parameters by a least
squares fit method. We have performed extensive Monte Carlo simulations by analyzing artificial signals added to white Gaussian noise. We have studied biases and variances of the estimators.
We find that our estimators, above a certain signal-to-noise ratio which is around 8,
are to a good accuracy unbiased and their variances are close to the ones calculated from the inverse
of the Fisher matrix.

\section*{Acknowledgments}

We would like to thank members of the LSC-Virgo CW data analysis group for
helpful discussions. This work was supported in part by the Polish Ministry of
Science and Higher Education grant DPN/N176/VIRGO/2009 and the National Science
Center grant UMO-2013/01/ASPERA/ST9/00001.


\newcommand{\PR}{Phys.\ Rev.\ }
\newcommand{\PRL}{Phys.\ Rev.\ Lett.\ }
\newcommand{\CQG}{Class.\ Quantum Grav.\ }
\newcommand{\mnras}{MNRAS}

\end{document}